\documentclass[prc,superscriptaddress,noshowpacs,unsortedaddress,twocolumn,showpacs,preprintnumbers,amsmath,amssymb]{revtex4-1}

\usepackage[dvipdfmx]{graphicx}
\usepackage{amsmath,amssymb,times}
\usepackage{color}
\usepackage{ulem}
\usepackage{bm}
\usepackage{here}

\def\lsim{~\,\makebox(1,1){$\stackrel{<}{\widetilde{}}$}\,~}

\newcommand{\beq}{\begin{equation}}
\newcommand{\eeq}{\end{equation}}
\newcommand{\bea}{\begin{eqnarray}}
\newcommand{\eea}{\end{eqnarray}}

\newcommand{\bfi}[1]{\mbox{\boldmath $#1$}}

\newcommand{\vK}{{\bfi K}}

\newcommand{\vs}{{\bfi s}}

\newcommand{\vrr}{{\bfi r}}
\newcommand{\vR}{{\bfi R}}

\def\a{\alpha}

\begin{document}

\title{Skin values and matter radii of  $^{208}$Pb and $^{58,60,64}$Ni\\
based on reaction cross section of $^{3,4}$He scattering}

\author{Shingo~Tagami}
%\affiliation{Department of Physics, Kyushu University, Fukuoka 819-0395, Japan}
%\email[]{sh.tagami@gmail.com}

\author{Tomotsugu~Wakasa}
%\affiliation{Department of Physics, Kyushu University, Fukuoka 819-0395, Japan}
%\email[]{wakasa@phys.kyushu-u.ac.jp} 

\author{Masanobu Yahiro}
%\email[]{orion093g@gmail.com}
%\affiliation{Department of Physics, Kyushu University, Fukuoka 819-0395, Japan}             

\begin{abstract}
\begin{description}
\item[Background]
The PREX group  reported a new skin value, 
$r_{\rm skin}^{208}({\rm PREX2}) = 0.283\pm 0.071= 0.212 \sim 0.354
\,{\rm fm}$. 
Using the chiral (Kyushu) $g$-matrix folding model with the proton and neutron densities determined with D1S+GHFB+AMP, we determined 
neutron skin thickness $r_{\rm skin}^{208}({\rm exp})=0.278 \pm 0.035$~fm from reaction cross sections  $\sigma_{\rm R}({\rm exp})$ of p+$^{208}$Pb scattering, where D1S-GHFB+AMP denotes  
Gogny-D1S HFB with angular momentum projection (AMP). 
The method also yielded $r_{\rm skin}^{208}({\rm exp}) = 0.416 \pm 0.146$ from  
$\sigma_{\rm R}({\rm exp})$ of $^{4}$He+$^{208}$Pb scattering. 
In our previous work, we accumulated the 206 EoSs and determined a sloop parameter 
$L=76 \sim 165$~MeV from the 206 EoSs. The value of $L$ yields  
$r_{\rm skin}^{208}=0.102 \sim 0.354~{\rm fm}$.  
As for  $^{58}$Ni, a novel method for measuring nuclear reactions in inverse kinematics with stored ion beams was successfully used to extract matter radii $r_{\rm m}$ of $^{58}$Ni at the GSI facility. 
\item[Purpose]
 As the first aim, we first determine  $r_{\rm skin}^{208}({\rm exp})$  
from $\sigma_{\rm R}({\rm exp})$ of  $^{3}$He scattering on $^{208}$Pb target and 
take the weighted mean and its error for $r_{\rm skin}^{208}({\rm PREX2})$, three skin values 
of p+$^{208}$Pb, $^{3, 4}$He+$^{208}$Pb scattering and 
$r_{\rm skin}^{208}=0.102 \sim 0.354~{\rm fm}$ based on the 206 EoSs.
As the second aim, we determine matter radii $r_{m}({\rm exp})$ of $^{58,60,64}$Ni  
from $\sigma_{\rm R}({\rm exp})$ of  $^{3,4}$He scattering on $^{58,60,64}$Ni targets.  
\item[Results]
Our result is  $r_{\rm skin}^{208}({\rm exp}) =0.512 \pm 0.268~{\rm fm}$ for  
$^{3}$He+$^{208}$Pb scattering.  
Our results based on  $^{4}$He+$^{58,60,64}$Ni scattering are 
$r_{\rm m}^{58}({\rm exp})=3.734 \pm	0.091$~fm,
$r_{\rm m}^{60}({\rm exp})=3.815 \pm	0.094$~fm,
$r_{\rm m}^{64}({\rm exp})=4.001 \pm	0.093$~fm.
Our results based on  $^{3}$He+$^{58,60,64}$Ni scattering are 
$r_{\rm m}^{58}({\rm exp})=3.709 \pm	0.098 $~fm,
$r_{\rm m}^{60}({\rm exp})=3.753 \pm	0.112$~fm,
$r_{\rm m}^{64}({\rm exp})=3.822 \pm	0.142$~fm. 
\item[Conclusion]
Our conclusion is  $r_{\rm skin}^{208} =0.285 \pm 0.030~{\rm fm}$. It is determined from 
the 5  skin values mentioned above.
\end{description}
 \end{abstract}

\maketitle

\section{Introduction and Conclusion}
\label{Sec:Introduction}

{\it Background:} 

Horowitz, Pollock and Souder proposed a direct measurement 
for neutron skin thickness $r_{\rm skin}=r_{\rm n}-r_{\rm p}$~\cite{PRC.63.025501}, 
where $r_{\rm p}$ and $r_{\rm n}$ are proton and neutron radii, respectively.  
This direct measurement consists of parity-violating and elastic electron scattering. 
In fact, the PREX collaboration has reported, 
\begin{equation}
r_{\rm skin}^{208}({\rm PREX2}) = 0.283\pm 0.071= 0.212 \sim 0.354
\,{\rm fm}, 
\label{Eq:Experimental constraint 208}
\end{equation}
combining the original Lead Radius EXperiment (PREX)  result \cite{PRL.108.112502,PRC.85.032501} 
with the updated PREX2 result \cite{Adhikari:2021phr}. 
This measurement is a direct one yielding  a neutron skin thickness $r_{\rm skin}^{208}$. 
The $r_{\rm skin}^{208}({\rm PREX2})$ is determined with weak electromagnetic interaction.

Many theoretical  predictions on the symmetry energy $S_{\rm sym}(\rho)$ have been made so far. 
In neutron star, the $S_{\rm sym}(\rho)$ and its density ($\rho$) dependence influence strongly the nature within the star.
The symmetry energy $S_{\rm sym}(\rho)$ cannot be measured by experiment directly. 
In place of  $S_{\rm sym}(\rho)$, the neutron-skin thickness $r_{\rm skin}$ is measured to determine 
the slope parameter $L$, since a strong correlation between $r_{\rm skin}^{208}$ and $L$ is well known~\cite{RocaMaza:2011pm}. In fact, the relation of Ref.~\cite{RocaMaza:2011pm} yields 
$L=76 \sim 172$~MeV from $r_{\rm skin}^{208}({\rm PREX2})$. 
As shown later in Eq.~\eqref{Eq:skin-const}, we also determine the relation yielding 
$L=76 \sim 165$~MeV.

Meanwhile, reaction cross section $\sigma_{\rm R}$ is a standard observable
for determining matter radius $r_{\rm m}$ and skin value $r_{\rm skin}$. 
High-accuracy data $\sigma_{\rm R}({\rm exp})$ with 2\% error are available for 
p + $^{208}$Pb scattering in $21  \leq E_{\rm lab} \leq 100$~MeV~\cite{PRC.12.1167,NPA.653.341,PRC.71.064606}. 
The chiral (Kyushu) $g$-matrix folding model with the D1S-GHFB+AMP  yields 
$r_{\rm skin}^{208}({\rm exp})=0.278 \pm 0.035$~fm~\cite{Tagami:2020bee}, where D1S-GHFB+AMP stands for 
Gogny-D1S HFB with angular momentum projection (AMP).  
The value is  consistent with   $r_{\rm skin}^{208}({\rm PREX2})$. Data $\sigma_{\rm R}({\rm exp})$ 
 with 5\% error are available for  $^{4}$He+$^{208}$Pb scattering~\cite{Ingemarsson:2000vfz}.
 We then determined   $r_{\rm skin}^{208}({\rm exp}) = 0.416 \pm 0.146$~fm in $29  \leq E_{\rm lab} \leq 48$~MeV 
 per nucleon~\cite{Matsuzaki:2021hdm}. 
In this paper, we firstly  extract $r_{\rm skin}^{208}({\rm exp})$ from $^{3}$He+$^{208}$Pb scattering 
by using the Kyushu $g$-matrix folding model with the D1S-GHFB+AMP, 
although the data~\cite{,Ingemarsson:2000vfz}  on the $\sigma_{\rm R}$ have large 8.1\% error. 

As for  $^{58}$Ni, a novel method for measuring nuclear reactions in inverse kinematics with stored ion beams was successfully used to extract the $r_{\rm m}$ of $^{58}$Ni~\cite{Zamora:2017adt}. The experiment was performed at the experimental heavy-ion storage ring at the GSI facility. 
Their result determined from the differential cross section 
for $^{58}$Ni+$^{4}$He scattering is $r_m({\rm GSI})=3.70(7)$fm for $^{58}$Ni.

Data $\sigma_{\rm R}$ with $4 \sim  7\%$ error are available 
for   $^{3,4}$He+$^{58,60,64}$Ni scattering~\cite{Ingemarsson:2000vfz}.

{\it   A theoretical constraint on $r_{\rm skin}^{208}$:}

\begin{figure}[H]
\begin{center}
 \includegraphics[width=0.5\textwidth,clip]{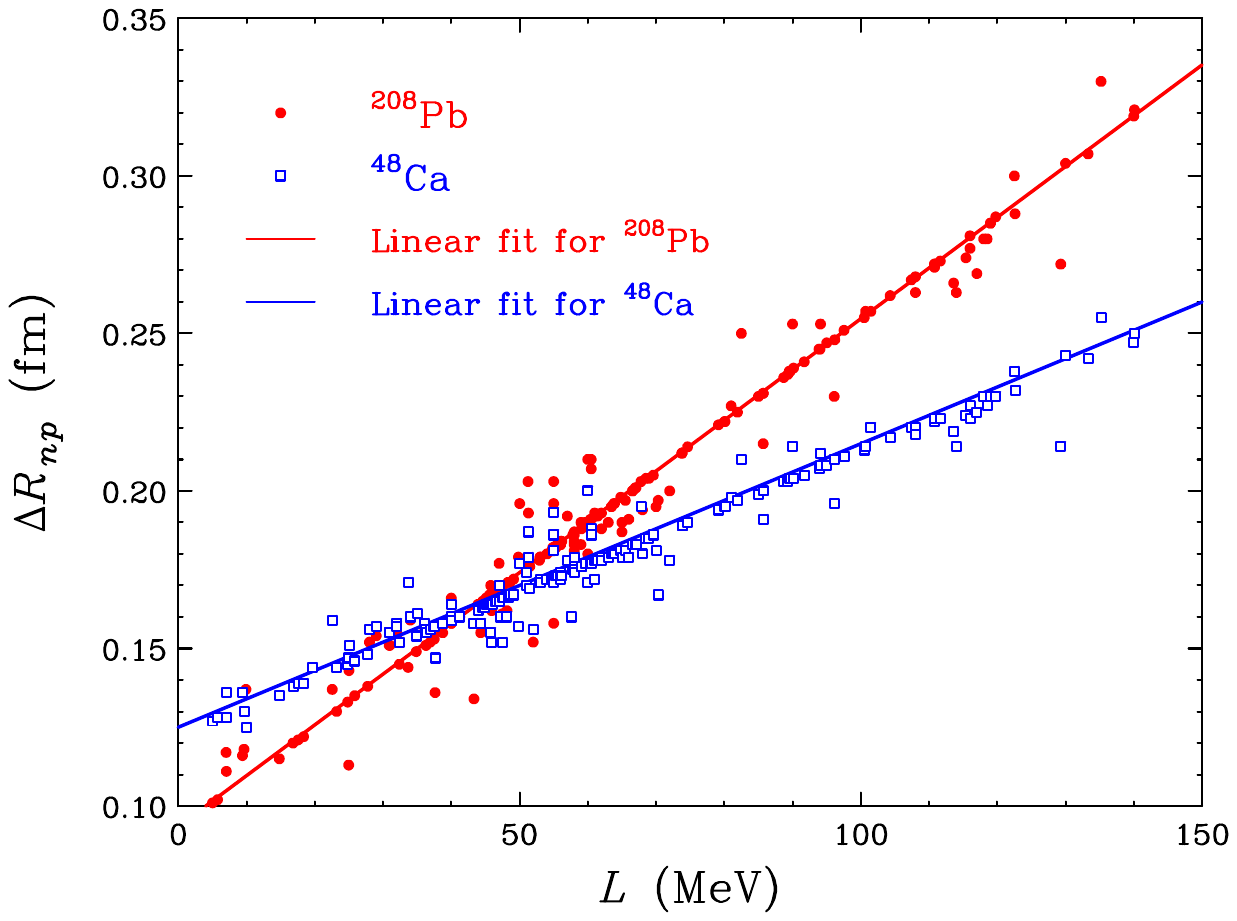}
 \caption{ 
Skin values $\Delta R_{np} \equiv r_{\rm skin}$ as a function of $L$ for $^{208}$Pb and  $^{48}$Ca.    
The straight line shows Eq.~\eqref{Eq:skin-L} for $^{208}$Pb and another line is 
$r_{\rm skin}^{48}=0.0009 L + 0.125$ for $^{48}$Ca.  
Dots denote  206 EoSs for $^{208}$Pb and $^{48}$Ca. 
   }
 \label{Fig-L-skin}
\end{center}
\end{figure}

In Ref.~\cite {TAGAMI2022105155}, we  accumulated the 206 EoSs from Refs.~\cite{Akmal:1998cf,RocaMaza:2011pm,Ishizuka:2014jsa,Gonzalez-Boquera:2017rzy,D1P-1999,Gonzalez-Boquera:2017uep,Oertel:2016bki,Piekarewicz:2007dx,Lim:2013tqa,Sellahewa:2014nia,Inakura:2015cla,Fattoyev:2013yaa,Steiner:2004fi,Centelles:2010qh,Dutra:2012mb,Brown:2013pwa,Brown:2000pd,Reinhard:2016sce,Tsang:2019ymt,Ducoin:2010as,Fortin:2016hny,Chen:2010qx,Zhao:2016ujh,Zhang:2017hvh,Wang:2014rva,Lourenco:2020qft} 
 in which $r_{\rm skin}^{208}$ and/or $L$ is presented, since 
a strong correlation between $r_{\rm skin}^{208}$ and a slope parameter $L$ is shown. 
The 206 EoSs are shown in Table I of Ref.~\cite {TAGAMI2022105155}.  
The correlation between $L$ and  $r_{\rm skin}^{208}$ is more reliable when the number of EoSs is larger. 
For this reason,  we  take  the 206 EoSs. 
Using the 206 EoSs of Table I, we found the $L$-$r_{\rm skin}^{208}$  relation as 
\bea
L{(r_{\rm skin}^{208})}=620.39~r_{\rm skin}^{208}-57.963 
\label{Eq:skin-L}
\eea
with 
a very high correlation coefficient $R=0.99$~\cite{TAGAMI2022105155}; see Fig. \ref{Fig-L-skin}. Note that a very high correlation coefficient means that 
Eq.~\eqref{Eq:skin-L} is concrete.

%%%%%%%%%%%%%%%%%%%%%%%
%%%  Figure
%%%%%%%%%%%%%%%%%%%%%%%t

The 206 EoSs yields a theoretical constraint on $L$, i.e., $L=5.020 \sim 161.05$~MeV.  
Using Eq.~\eqref{Eq:skin-L} and $L=5.020 \sim 161.05$~MeV, we can derive a range of $r_{\rm skin}^{208}$ theoretically. 
The result is  
\bea
 r_{\rm skin}^{208}=0.102 \sim 0.354~{\rm fm} .
\label{Eq:skin-const}
\eea
This is a theoretical constraint on $r_{\rm skin}^{208}$.   The upper bound of the theoretical constraint 
agrees with that of $r_{\rm skin}^{208}({\rm PREX2})$. 
The $r_{\rm skin}^{208}({\rm PREX2})$ supports stiffer EoSs.

%%%%%%%%%%%%%%%%%%%%%%%
%%%  Figure
%%%%%%%%%%%%%%%%%%%%%%%t
\begin{figure}[H]
\begin{center}
 \includegraphics[width=0.5\textwidth,clip]{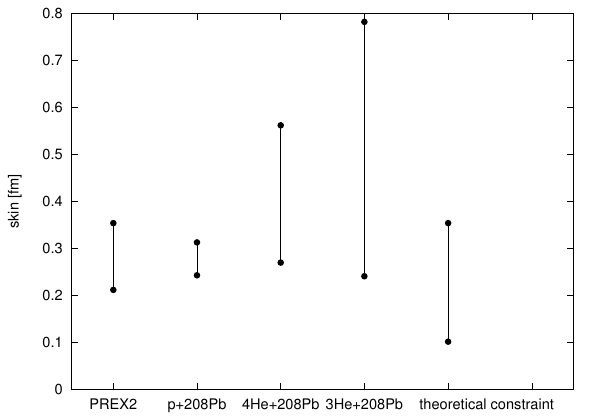}
 \caption{ 
 $r_{\rm skin}^{208}({\rm PREX2}) $, three skin values  determined from 
 reaction cross sections $\sigma_{\rm R}$  of p.  :$^{4}$He, $^{3}$He scattering on 
 $^{208}$Pb target, the theoretical constraint of Eq.~\eqref{Eq:skin-const}.
   }
 \label{Fig-skins}
\end{center}
\end{figure}

As for  $r_{\rm skin}^{208}$, therefore, we can summarize five skin-values; 1)~PREX2, 
2)~p scattering, 3)~$^{4}$He scattering, 4)~$^{3}$He scattering, 5)~the theoretical constraint 
of Eq.~\eqref{Eq:skin-const} in Fig.  \ref{Fig-skins}. 
The skin value $r_{\rm skin}^{208}({\rm exp}) $ of $^{3}$He+$^{208}$Pb scattering 
supports stiffer EoSs.

{\it Aims:}
 
 As the first aim, we determine $r_{\rm skin}^{208}({\rm exp})$   
from the $\sigma_{\rm R}({\rm exp})$ of  $^{3}$He+$^{208}$Pb scattering  and 
take the weighted mean and its error for five skin values  1), 2), 3), 4), 5) 
shown in Fig.  \ref{Fig-skins}.
As the second aim, we determine $r_{m}({\rm exp})$ of $^{58,60,64}$Ni  
from the $\sigma_{\rm R}({\rm exp})$ of  $^{3,4}$He scattering on $^{58,60,64}$Ni targets.

{\it Results:}

 We first determine  $ r_{\rm skin}^{208}({\rm exp})$ from $\sigma_{\rm R}(\rm exp)$~\cite{Ingemarsson:2000vfz}  of 
$^{3}$He+ $^{208}$Pb scattering in  $32 \leq E_{\rm lab} \leq 56$~MeV 
 per nucleon; the result is    
 \bea
 r_{\rm skin}^{208}({\rm exp}) =0.512 \pm 0.268=0.243 \pm 0.780~{\rm fm} , 
 \label{Eq:skin-Pb-He3}
 \eea
where $E_{\rm lab}$ is an incident energy per nucleon. 
 The result is consistent with $r_{\rm skin}^{208}({\rm PREX2}) = 0.212 \sim 0.354$~fm.

We next evaluate $r_{\rm m}^{58,60,64}({\rm exp})$ from $^{3,4}$He+$^{58,60,64}$Ni scattering. 
Our results for $^{4}$He scattering are 
$r_{\rm m}^{58}({\rm exp})=3.734 \pm	0.091$~fm,
$r_{\rm m}^{60}({\rm exp})=3.815 \pm	0.094$~fm,
$r_{\rm m}^{64}({\rm exp})=4.001 \pm	0.093$~fm, where 
the data taken is Ref.~\cite{Ingemarsson:2000vfz} for $^{58,60}$Ni
and is Ref.~\cite{D-He4+Ni64} for $^{64}$Ni.
Our results for  $^{3}$He scattering are 
$r_{\rm m}^{58}({\rm exp})=3.709 \pm	0.098 $~fm,
$r_{\rm m}^{60}({\rm exp})=3.753 \pm	0.112$~fm,
$r_{\rm m}^{64}({\rm exp})=3.822 \pm	0.142$~fm, 
where 
the data taken is Ref.~\cite{Ingemarsson:2000vfz} for $^{58,60,64}$Ni.
Our results  based on  $^{4}$He+$^{58,60,64}$Ni are consistent with those based on $^{3}$He+$^{58,60,64}$Ni. 
In addition, these results for $^{58}$Ni are consistent with the $r_{\rm m}$ of Ref.~\cite{Zamora:2017adt}.

{\it Conclusion:}
We  take the weighted mean and its error for 
five skin values  1), 2), 3), 4), 5) shown in Fig.  \ref{Fig-skins}.
The skin value is 
\bea
r_{\rm skin}^{208} =0.285 \pm 0.030~{\rm fm}.
\eea
This value is mainly determined by the skin value of  p+$^{208}$Pb scattering.

\section{Mehod}
\label{Sec:Method}

Our method is composed of the chiral (Kyushu) $g$-matrix  folding model for scattering~\cite{Toyokawa:2017pdd} 
and D1S-GHFB+AMP~\cite{Tagami:2019svt} for target densities, where D1S-GHFB+AMP stands for 
Gogny-D1S HFB with the angular momentum projection (AMP). 

As for the symmetric nuclear matter, Kohno calculated the $g$ matrix 
by using the Brueckner-Hartree-Fock (BHF) method with chiral N$^{3}$LO 2NFs and NNLO 3NFs~\cite{Kohno:2012vj}. 
The framework is applied for positive energies. The resulting non-local chiral  $g$ matrix is localized 
into three-range Gaussian forms by using the localization method proposed 
by the Melbourne group~\cite{von-Geramb-1991,Amos-1994}. 
The resulting local  $g$ matrix is referred to as Kyushu  
$g$-matrix in this paper~\cite{Toyokawa:2017pdd}.

The Kyushu $g$-matrix folding model was tested for $^{12}$C scattering 
on  $^{9}$Be, $^{12}$C, $^{27}$Al targets in  $30  \lsim E_{\rm lab} \lsim 400 $~MeV. 
We found that the Kyushu $g$-matrix folding model is good 
in $30  \lsim E_{\rm lab} \lsim 100 $~MeV and $250  \lsim E_{\rm lab} \lsim 400$~MeV.

The brief formulation of the folding model itself is shown below. 
The potential $U$ consists 
of the direct part ($U^{\rm DR}$) and the exchange part ($U^{\rm EX}$):
\bea
\label{eq:UD}
U^{\rm DR}(\vR) \hspace*{-0.15cm} &=& \hspace*{-0.15cm} 
\sum_{\mu,\nu}\int \rho^{\mu}_{\rm P}(\vrr_{\rm P}) 
            \rho^{\nu}_{\rm T}(\vrr_{\rm T})
            g^{\rm DR}_{\mu\nu}(s;\rho_{\mu\nu}) d \vrr_{\rm P} d \vrr_{\rm T}, \\
\label{eq:UEX}
U^{\rm EX}(\vR) \hspace*{-0.15cm} &=& \hspace*{-0.15cm}\sum_{\mu,\nu} 
\int \rho^{\mu}_{\rm P}(\vrr_{\rm P},\vrr_{\rm P}-\vs)
\rho^{\nu}_{\rm T}(\vrr_{\rm T},\vrr_{\rm T}+\vs) \nonumber \\
            &&~~\hspace*{-0.5cm}\times g^{\rm EX}_{\mu\nu}(s;\rho_{\mu\nu}) \exp{[-i\vK(\vR) \cdot \vs/M]}
            d \vrr_{\rm P} d \vrr_{\rm T},~~~~
            \label{U-EX}
\eea
where $\vs=\vrr_{\rm P}-\vrr_{\rm T}+\vR$ 
for the coordinate $\vR$ between P and T. The coordinate 
$\vrr_{\rm P}$ 
($\vrr_{\rm T}$) denotes the location for the interacting nucleon 
measured from the center-of-mass of the projectile (target). 
Each of $\mu$ and $\nu$ stands for the $z$-component
of isospin; 1/2 means neutron and $-$1/2 does proton.
The original form of $U^{\rm EX}$ is a non-local function of $\vR$,
but  it has been localized in Eq.~\eqref{U-EX}
with the local semi-classical approximation~\cite{Brieva-Rook-1,Brieva-Rook-2,Brieva-Rook-3} in which
P is assumed to propagate as a plane wave with
the local momentum $\hbar \vK(\vR)$ within a short range of the 
nucleon-nucleon interaction, where $M=A A_{\rm T}/(A +A_{\rm T})$
for the mass number $A$ ($A_{\rm T}$) of P (T).
The validity of this localization is shown in Ref.~\cite{Minomo:2009ds}.

The direct and exchange parts, $g^{\rm DR}_{\mu\nu}$ and 
$g^{\rm EX}_{\mu\nu}$, of the effective nucleon-nucleon interaction 
($g$-matrix) are assumed to depend on the local density
\bea
 \rho_{\mu\nu}=\sigma^{\mu} \rho^{\nu}_{\rm T}(\vrr_{\rm T}+\vs/2)
\label{local-density approximation}
\eea
at the midpoint of the interacting nucleon pair, where $\sigma^{\mu}$ is the Pauli matrix of a nucleon in P. 
This choice of  the local density is quite successful for $^{4}$He scattering, as shown in Ref. \cite{PRC.89.064611}. 

The direct and exchange parts are described by
\begin{align}
&\hspace*{0.5cm} g_{\mu\nu}^{\rm DR}(s;\rho_{\mu\nu}) \nonumber \\ 
&=
\begin{cases}
\displaystyle{\frac{1}{4} \sum_S} \hat{S}^2 g_{\mu\nu}^{S1}
 (s;\rho_{\mu\nu}) \hspace*{0.42cm} ; \hspace*{0.2cm} 
 {\rm for} \hspace*{0.1cm} \mu+\nu = \pm 1 
 \vspace*{0.2cm}\\
\displaystyle{\frac{1}{8} \sum_{S,T}} 
\hat{S}^2 g_{\mu\nu}^{ST}(s;\rho_{\mu\nu}), 
\hspace*{0.2cm} ; \hspace*{0.2cm} 
{\rm for} \hspace*{0.1cm} \mu+\nu = 0 
\end{cases}
\\
&\hspace*{0.5cm}
g_{\mu\nu}^{\rm EX}(s;\rho_{\mu\nu}) \nonumber \\
&=
\begin{cases}
\displaystyle{\frac{1}{4} \sum_S} (-1)^{S+1} 
\hat{S}^2 g_{\mu\nu}^{S1} (s;\rho_{\mu\nu}) 
\hspace*{0.34cm} ; \hspace*{0.2cm} 
{\rm for} \hspace*{0.1cm} \mu+\nu = \pm 1 \vspace*{0.2cm}\\
\displaystyle{\frac{1}{8} \sum_{S,T}} (-1)^{S+T} 
\hat{S}^2 g_{\mu\nu}^{ST}(s;\rho_{\mu\nu}) 
\hspace*{0.2cm} ; \hspace*{0.2cm}
{\rm for} \hspace*{0.1cm} \mu+\nu = 0 ~~~~~
\end{cases}
\end{align}
where $\hat{S} = {\sqrt {2S+1}}$ and $g_{\mu\nu}^{ST}$ are 
the spin-isospin components of the $g$-matrix interaction.

The proton and neutron densities, $\rho_{\rm p}(r)$ and $\rho_{\rm n}(r)$, are calculated with 
D1S-GHFB+AMP~\cite{Tagami:2019svt}. 
 As a way of taking the center-of-mass correction to the D1S-GHFB+AMP densities, 
we use the method of Ref.~\cite{Sumi:2012fr}, since the procedure is quite simple. 
For $^{3,4}$He, we take the phenomenological densities of Ref.~\cite{C12-density}.

In order to deduce the $r_{\rm m}$ from  $\sigma_{\rm R}({\rm exp})$~\cite{Bonin:1985rr,Ingemarsson:2000vfz}, 
we have to scale the proton and neutron densities, as shown in Sec.~\ref{Results}. 
Now we explain the scaling of density $\rho(\vrr)$.  
We can obtain the scaled density $\rho_{\rm scaling}(\vrr)$ from the original density $\rho(\vrr)$ as
\bea
\rho_{\rm scaling}(\vrr)=\frac{1}{\a^3}\rho(\vrr/\a)
\eea
with a scaling factor
\bea
\a=\sqrt{ \frac{\langle \vrr^2 \rangle_{\rm scaling}}{\langle \vrr^2 \rangle}} .\eea
As for $^{208}$Pb, $\a_p=1$ and  $\a_n=1.067341$ are close to 1.

%Results
\section{Results}
\label{Results}

We first determine $r_n({\rm PREX2})=5.727 \pm 0.071$ fm and $r_m({\rm PREX2})=5.617 \pm 0.044$ fm 
from $r_p({\rm exp})$ = 5.444 fm \cite{Brown:2013mga} and $r_{\rm skin}^{208}({\rm PREX2})$. 
The $r_p({\rm GHFB})$  calculated with D1S-GHFB+AMP agrees with 
$r_p({\rm exp})$= 5.444 fm of electron scattering. 
For simplicitly, we refer the D1S-GHFB+AMP proton density to as the PREX2 proton density 
in this paper.   
The neutron density calculated with GHFB+AMP is scaled so as to  $r_n({\rm PREX2})=5.727$ fm. 
The resulting neutron density is referred to as the neutron PREX2 density in this paper.

\begin{figure}[H]
\begin{center}
 \includegraphics[width=0.5\textwidth,clip]{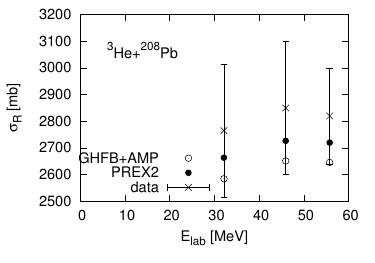}
 \caption{ 
 $E_{\rm lab}$ dependence of reaction cross sections $\sigma_{\rm R}$ 
 for $^{3}$He+$^{208}$Pb scattering. 
 Open circles stand for the results of the  Kyushu $G$-matrix folding  model with the 
 D1S-GHFB+AMP densities. 
 Closed circles correspond to  the PREX2 proton and neutron densities. 
 The data are taken from Refs.~\cite{Ingemarsson:2000vfz}. 
   }
 \label{Fig-RXsec-He3+Pb-1}
\end{center}
\end{figure}

Figure~\ref{Fig-RXsec-He3+Pb-1} shows $E_{\rm lab}$ of $\sigma_{\rm R}$ for 
$^{3}$He+$^{208}$Pb scattering in  $32 \leq E_{\rm lab} \leq 56$~MeV per nucleon.
The Kyushu $g$-matrix folding model with D1S-GHFB+AMP 
undershoots  $\sigma_{\rm R}({\rm exp})$~\cite{Ingemarsson:2000vfz} 
in $32 \leq E_{\rm lab} \leq 56$~MeV by a factor of 0.9344. 
This indicates that our results are good enough.
The  Kyushu $g$-matrix folding  model with the PREX2 neutron and proton 
densities reproduces 
the data on $\sigma_R$ at $E_{\rm lab} = 32, 46, 56$ MeV per nucleon.

%%%%%%%%%%%%%%%%%%%%%%%
%%%  Figure
%%%%%%%%%%%%%%%%%%%%%%%

In $32 \leq E_{\rm lab} \leq 56$~MeV per nucleon, 
we can obtain  $r_{\rm m}({\rm exp})$ from $\sigma_{\rm R}({\rm exp})$ 
by scaling the D1S-GHFB+AMP densities so as to reproduce $\sigma_{\rm R}({\rm exp})$ for each $E_{\rm lab}$ 
under the condition that the proton radius of the scaling density agrees with the data 
$r_{\rm p}({\rm exp})=5.444$~{\rm fm}~\cite{Brown:2013mga} of electron scattering,  
and take the weighted mean and its error for the resulting $r_{\rm m}({\rm exp})$. 
From the resulting $r_{m}({\rm exp})=5.759 \pm 0.169$~fm and 
$r_{\rm p}({\rm exp})=5.444$~{\rm fm} of electron scattering, 
we can get $r_{\rm skin}^{208}({\rm exp})=0.512 \pm 0.268$~fm and 
$r_{\rm n}^{208}({\rm exp})=5.956 \pm 0.268$~fm.

The same procedure is taken for $^{3,4}$He+$^{58,60,64}$Ni scattering. 
As for $^{4}$He, we take the data~\cite{Ingemarsson:2000vfz}  in $30 \leq E_{\rm lab} \leq 48$~MeV per nucleon 
for $^{58,60}$Ni and the data~\cite{D-He4+Ni64} in $E_{\rm lab}=8.8$~MeV per nucleon for $^{64}$Ni.  
As for $^{3}$He, we take the data~\cite{Ingemarsson:2000vfz}  in $32 \leq E_{\rm lab} \leq 56$~MeV per nucleon 
for $^{58,60}$Ni and the data~\cite{Ingemarsson:2000vfz} in $E_{\rm lab}=32$~MeV per nucleon for $^{64}$Ni. 
Our results are shown in Sec. \ref{Sec:Introduction}.

\section*{Acknowledgements}
We thank Dr. Toyokawa for his contribution.

\bibliography{Folding-v11}

\end{document}